\documentstyle[12pt]{article}
\newcommand{\bea}{\begin{eqnarray}}
\newcommand{\eea}{\end{eqnarray}}
\newcommand{\be}{\begin{equation}}
\newcommand{\ee}{\end{equation}}
\newcommand{\vs}[1]{\vspace{#1 mm}}

\renewcommand{\a}{\alpha}
\renewcommand{\b}{\beta}

\renewcommand{\d}{\delta}

\newcommand{\dsl}{\pa \kern-0.5em /}

\newcommand{\pa}{\partial}

\newcommand{\nn}{\nonumber\\}

\begin{document}
\topmargin 0pt
\oddsidemargin 0mm



\begin{flushright}

USTC-ICTS-06-05\\

hep-th/0610264\\


\end{flushright}

\vspace{2mm}

\begin{center}

{\Large \bf Intersecting non-SUSY $p$-brane with chargeless 0-brane
as black $p$-brane

}

\vs{6}

{\large Hua Bai$^a$\footnote{E-mail: huabai@itp.ac.cn},
J. X. Lu$^b$\footnote{E-mail: jxlu@ustc.edu.cn}
 and Shibaji Roy$^c$\footnote{E-mail: shibaji.roy@saha.ac.in}}

 \vspace{4mm}

{\em
$^a$ Institute of Theoretical Physics, Chinese Academy of Sciences,\\

P.O.Box 2735, Beijing 100080, China\\

\vs{4}

 $^b$ Interdisciplinary Center for Theoretical Study\\

 University of Science and Technology of China, Hefei, Anhui
 230026, China\\




\vs{4}

 $^c$ Saha Institute of Nuclear Physics,
 1/AF Bidhannagar, Calcutta-700 064, India}

\end{center}

\begin{abstract}

Unlike BPS $p$-brane, non-supersymmetric (non-susy) $p$-brane could
be either charged or chargeless. As envisaged in [hep-th/0503007],
we construct an intersecting non-susy $p$-brane with chargeless
non-susy $q$-brane by taking T-dualities along the delocalized
directions of the non-susy $q$-brane solution delocalized in $(p-q)$
transverse directions (where $p\geq q$). In general these solutions
are characterized by four independent parameters. We show that when
$q=0$ the intersecting charged as well as chargeless non-susy
$p$-brane with chargeless 0-brane can be mapped by a coordinate
transformation to black $p$-brane when two of the four parameters
characterizing the solution take some special values. For
definiteness we restrict our discussion to space-time dimensions
$d=10$. We observe that parameters characterizing the black brane
and the related dynamics are in general in a different branch of the
parameter space from those describing the brane-antibrane
annihilation process. We demonstrate this in the two examples,
namely, the non-susy D0-brane and the intersecting non-susy D4 and
D0-branes, where the solutions with the explicit microscopic
descriptions are known.

\end{abstract}
\newpage
\section{Introduction}

The static, non-susy and asymptotically flat $p$-brane
solutions\footnote{We use the terminology non-susy $p$-brane to
represent generically either the $p$-brane-anti$p$-brane system or
the non-BPS $p$-branes.} having isometries ISO($p,1$) $\times$
SO($d-p-1$) of type II supergravities in arbitrary space-time
dimensions ($d$) are given in ref.\cite{Lu:2004ms}\footnote{See
\cite{Zhou:1999nm,Ivashchuk:2001ra} for earlier works on non-susy
$p$-branes.}. Unlike the BPS $p$-branes characterized by a single
unknown parameter, these solutions are characterized by three
unknown parameters and could be either charged or chargeless with
respect to a $(p+1)$-form gauge field. It is well-known that these
non-susy $p$-branes have a natural interpretation as coincident
$p$-brane-anti-$p$-brane (or non-BPS $p$-brane)
\cite{Brax:2000cf,Lu:2004dp,Bai:2005jr}. Since there is an open
string tachyon on their world-volume, these systems are unstable
\cite{Sen:1999mg}. The three parameters in the supergravity
solutions then have physical meanings in terms of number of branes,
number of antibranes and the tachyon parameter. The explicit
semi-empirical relations of the supergravity parameters in terms of
the microscopic physical parameters are given in
ref.\cite{Lu:2004dp,Bai:2005jr}, so that it correctly captures the
picture of tachyon condensation for these systems.

On the other hand, there is another class of non-susy $p$-brane
solutions in type II supergravities in the form of black $p$-branes
with isometries R $\times$ ISO($p$) $\times$ SO($d-p-1$)
\cite{Horowitz:1991cd,Duff:1993ye,Duff:1994an}. Unlike the previous
solutions which can have naked singularities for certain choices of
parameters, the black $p$-branes always have regular horizons and
are characterized by two parameters corresponding to the mass and
the charge of the black branes. In the literature these two classes
are referred to as type 1 and type 2 solutions \cite{Lu:1996er}.

However, in contrary to the previous claim \cite{Lu:1996er}, we will argue that
there is only one class of non-susy $p$-brane solutions, namely,
the type 1.  We will restrict our discussion to $d=10$ for
definiteness. The black brane solutions, i.e. the type 2, with the
above isometries belong to a more general non-susy $p$-brane
solutions of type 1 than the one mentioned in the first paragraph.
We will show that the black $p$-brane solutions are nothing but
the intersecting\footnote{A particular kind of intersecting non-susy
brane solutions were constructed previously in \cite{Miao:2004bn,Bai:2005jr}.}
non-susy $p$-branes with the chargeless 0-branes
expressed in a different coordinate system, when two of the four
parameters characterizing the latter solutions take some special
values. In order to show that, we will first construct an
intersecting non-susy brane solutions where the constituent branes
are the charged non-susy $p$-brane and the chargeless non-susy
$q$-brane. The explicit solutions when both the non-susy branes in
the intersecting solution are charged are known only when $p-q=4$ \cite{Bai:2005jr}
(or the simple case $p - q = 0$ \cite{Lu:2004ms}). Here we are interested in
solutions when non-susy $q$-brane is chargeless. As mentioned in
ref.\cite{Lu:2005ju}, these solutions can be obtained from the non-susy
$q$-brane solutions delocalized in $(p-q)$ transverse directions
and then applying T-dualities along those delocalized directions.
We then make all the T-dual directions isometric resulting in an
intersecting non-susy $p$-brane with chargeless non-susy $q$-brane
solutions (where $p\geq q$). These solutions belong to type 1 and
as we will show are characterized by four independent parameters.
Next we show that for $q=0$, these solutions can be mapped to the
black $p$-brane solutions mentioned above as type 2, by a
coordinate transformation when two of the four parameters take
some special values. This possibility is expected since the
presence of 0-brane breaks the isometry of non-susy $p$-brane from
ISO($p,1$) $\times$ SO($9-p$) to R $\times$ ISO($p$) $\times$
SO($9-p$) and the black brane has two parameters not four as in
the non-susy brane solutions.

The above observation, however, gives rise to a puzzle. As we
mentioned for the non-susy $p$-branes, the intersecting non-susy
$p$-brane with a non-susy $q$-brane also is an unstable system which
is manifested by the presence of open string tachyon on their
world-volume. Since the parameters of these solutions are related to
the microscopic physical parameters (this is explicitly known for
$p-q=0, 4$) of the intersecting brane-antibrane system
\cite{Lu:2004dp,Bai:2005jr}, the SUGRA parameters change their
values as the brane-antibrane annihilation process (or the tachyon
condensation) takes place and finally reach their end values
associated with the BPS configuration. However, in this process we
never encounter the formation of a regular horizon for these
systems. Then how for $q=0$ and certain values of the parameters we
get a black brane solutions? As we will see, there actually exists
an additional disjoint branch\footnote{Note, however, that for the
complete chargeless case there is a meet point at the initial
tachyon value $T = 0$, i.e., at the top of the tachyon potential as
will be shown in the examples given later in this paper.} in the
parameter space which has been neglected so far. This branch can
give rise to the regular horizon formation for the black brane where
the underlying dynamics is governed by the possible closed string
tachyon condensation as discussed by Horowitz in
\cite{Horowitz:2005vp,Horowitz:2006mr} and evolves into a non-susy
``bubble of nothing''. In this branch one of the parameters of the
solution becomes unbounded in general and as such makes the solution
complex \cite{Lu:2004ms}, signalling a possible phase transition and
this is consistent with our above observation. We will give some
evidence for this. However, in this paper we limit ourselves to
finding the values of the parameters required for the black brane in
general only from the second branch and explain some related issues
leaving a detailed study along with other related issues to a future
publication. Also our discussion for providing evidences is limited
only to the case of non-susy D0-branes (which includes D0/${\bar
{\rm D}}0$ and non-BPS D0-branes) and intersecting non-susy
D4-branes with chargeless non-susy D0-branes. The reason for
choosing these two systems is that only for these two cases we have
the explicit relations between the SUGRA parameters and the
microscopic physical parameters describing correctly the process of
open string tachyon condensation \cite{Lu:2004dp,Bai:2005jr}.

This paper is organized as follows. In section 2, we give the four
parameter solution of intersecting non-susy $p$-brane with
chargeless non-susy $q$-brane. In section 3, we show that for $q=0$,
these solutions get mapped to black $p$-brane when two of the four
parameters take some specific values. In section 4, we use the
microscopic descriptions of D0/${\bar {\rm D}}0$ and also the
intersecting D4/${\bar {\rm D}}4$ and chargeless D0/${\bar {\rm
D}}0$ system to show that the parameter space characterizing the
usual black branes and the related dynamics is in a different branch
from that describing the brane-antibrane annihilation process ending
up with a BPS or Minkowski configuration. Our conclusion is
presented in section 5.

\section{Intersecting brane solution}

In this section we will give the construction of intersecting
non-susy $p$-brane with chargeless non-susy $q$-brane, where,
$p\geq q$. For this purpose we will start with the non-susy
$q$-brane solution delocalized in $(p-q)$ transverse directions
given in ref.\cite{Lu:2005jc}. The application of T-duality along each of the
$(p-q)$ delocalized directions successively will give an
anisotropic intersecting $p$ brane with chargeless
$q$-brane\footnote{Note that for BPS branes usually the same
procedure does not give intersecting solutions and we get an
isotropic, localized $p$-brane. The reason for getting a different
solution here is that the non-susy solutions involve more
parameters and break the isometry of the solution which can then
be regarded as intersecting solutions as mentioned in ref.\cite{Lu:2005ju}.}.
Then equating all the parameters associated with the various
delocalized directions we get the isotropic, intersecting
$p$-brane with chargeless $q$-brane. For $p-q=4$ the intersecting
solutions, when both the non-susy $p$-brane and the non-susy
$q$-brane are charged, are known explicitly \cite{Bai:2005jr} and we will compare
these solutions with the ones we obtain here. To begin let us
write below the non-susy $q$-brane solution delocalized in $(p-q)$
transverse directions given in eq.(5) of ref.\cite{Lu:2005jc} as, \bea ds^2 &=&
F^{\frac{q+1}{8}} (H{\tilde {H}})^{\frac{2}{7-p}}
\left(\frac{H}{\tilde
H}\right)^{(-2\sum_{i=2}^{p-q+1}\delta_i)/(7-p)}\left(dr^2 + r^2
d\Omega_{8-p}^2\right)\nn & & \qquad\qquad +
F^{-\frac{7-q}{8}}\left(-dt^2 + \sum_{i=1}^q (dx^i)^2 \right) +
F^{\frac{q+1}{8}} \sum_{i=2}^{p-q+1}\left(\frac{H}{\tilde
H}\right)^{2\d_i} (dx^{q+i-1})^2 \nn e^{2\phi} &=& F^{-a}
\left(\frac{H}{\tilde {H}}\right)^{2\delta_1},\qquad
F_{[8-q]}\,\,\, =\,\,\, b\,\,{\rm Vol}(\Omega_{8-p})\wedge
dx^{q+1} \ldots \wedge dx^{p} \eea Note that we have written the
solution (5) of ref.\cite{Lu:2005jc} for $d=10$ and we have replaced $p$ there
by $q$ and $q$ there by $(p-q)$ to represent non-susy $q$-brane
delocalized in $(p-q)$ directions. The various functions appearing
in (1) are defined below, \bea F &=& \left(\frac{H}{\tilde
H}\right)^\a \cosh^2\theta - \left(\frac{\tilde H}{H} \right)^\b
\sinh^2\theta\nn H &=& 1 + \frac{\omega^{7-p}}{r^{7-p}}\nn \tilde
H &=& 1 - \frac{\omega^{7-p}}{r^{7-p}} \eea where $H$ and $\tilde
H$ are two harmonic functions and $\a$, $\b$, $\theta$, $\d_1$,
$\d_2$, \ldots, $\d_{p-q+1}$, and $\omega$ are $(p-q+5)$
integration constants and $b$ is the charge parameter. However,
there are three relations among the parameters given as, \bea &
&\a-\b = a\d_1\nn & &\frac{1}{2} \d_1^2 + \frac{1}{2} \a (\a
-a\d_1) + \frac{2\sum_{i > j=2}^{p-q+1}\d_i\d_j} {7-p} =
\left(1-\sum_{i=2}^{p-q+1} \d_i^2\right) \frac{8-p}{7-p}\nn & & b
= (7-p) \omega^{7-p} (\a+\b)\sinh2\theta \eea where $a=(q-3)/2$.
So, using the above relations we can eliminate three of the
parameters and the solution therefore has $(p-q+3)$ independent
parameters. We will take T-duality along $x^{q+1}$, $x^{q+2}$,
\ldots, $x^p$. For that we write the metric in (1) in the string
frame as, \bea ds_{\rm str}^2 &=& e^{\phi/2} ds^2\nn &=&
F^{\frac{1}{2}} (H{\tilde {H}})^{\frac{2}{7-p}}
\left(\frac{H}{\tilde
H}\right)^{(-2\sum_{i=2}^{p-q+1}\delta_i)/(7-p) +
\frac{\d_1}{2}}\left(dr^2 + r^2 d\Omega_{8-p}^2\right)\nn & &  +
F^{-\frac{1}{2}} \left(\frac{H}{\tilde
H}\right)^{\frac{\d_1}{2}}\left(-dt^2 + \sum_{i=1}^q (dx^i)^2
\right) + F^{\frac{1}{2}} \sum_{i=2}^{p-q+1}\left(\frac{H}{\tilde
H}\right)^{2\d_i+\frac{\d_1}{2}} (dx^{q+i-1})^2 \eea Now when we
apply T-dualities \cite{Bergshoeff:1995as,Das:1996je} successively along
the delocalized directions
the dilaton changes as, \be e^{2\tilde \phi} = F^{\frac{3-p}{2}}
 \prod_{i=2}^{p-q+1}\left(\frac{H}{\tilde
H}\right)^{- 2\d_i} \left(\frac{H}{\tilde
H}\right)^{\frac{\d_1}{2}(4-p+q)} \ee T-dualities will not change
the various components of the string metric in (4) except the ones
$g_{q+1,q+1}$, $g_{q+2,q+2}$, \ldots, $g_{p,p}$. Let us write
below the changed components, \bea \tilde{g}^{\rm str}_{q+1,q+1}
&=& F^{-\frac{1}{2}}\left(\frac{H}{\tilde H}\right)^{-2\d_2 -
\frac{\d_1}{2}}\nn \tilde{g}^{\rm str}_{q+2,q+2} &=&
F^{-\frac{1}{2}}\left(\frac{H}{\tilde H}\right)^{-2\d_3 -
\frac{\d_1}{2}}\nn \quad\vdots & &\nn \tilde{g}^{\rm str}_{p,p}
&=& F^{-\frac{1}{2}}\left(\frac{H}{\tilde H}\right)^{-2\d_{p-q+1}
- \frac{\d_1}{2}} \eea Now we rewrite the full T-dual solution in
Einstein frame as \bea d\tilde{s}^2 &=& e^{-\tilde{\phi}/2}
d\tilde{s}_{\rm str}^2\nn &=& F^{\frac{p+1}{8}} (H{\tilde
{H}})^{\frac{2}{7-p}} \left(\frac{H}{\tilde
H}\right)^{\frac{(p-q)\d_1}{8} +
\frac{(3-p)\sum_{i=2}^{p-q+1}\delta_i}{2(7-p)}} \left(dr^2 + r^2
d\Omega_{8-p}^2\right)\nn & &  + F^{-\frac{7-p}{8}}
\left(\frac{H}{\tilde H}\right)^{\frac{(p-q)\d_1}{8} +
\sum_{i=2}^{p-q+1}\frac{\d_i}{2}}\left(-dt^2 + \sum_{i=1}^q
(dx^i)^2 \right)\nn & & + F^{-\frac{7-p}{8}}
\sum_{i=2}^{p-q+1}\left(\frac{H}{\tilde H}\right)^{-\d_1 +
\frac{\d_1}{8}(p-q) - \frac{3\d_i}{2} +\sum_{j(\neq i)=2}^{p-q+1}
\frac{\d_j}{2}} (dx^{q+i-1})^2\nn e^{2\tilde\phi} &=&
F^{\frac{3-p}{2}} \left(\frac{H}{\tilde
H}\right)^{\frac{\d_1}{2}(4-p+q) - \sum_{i=2}^{p-q+1} 2\d_i}\nn
F_{[8-p]} &=& b {\rm Vol}(\Omega_{8-p}) \eea where
$d\tilde{s}_{\rm str}^2$ is the metric in (4) with the
$g_{q+1,q+1}$, $g_{q+2,q+2}$, \ldots, $g_{p,p}$ replaced by the
ones given in eq.(6). Now we make all the delocalized directions
isotropic by setting $\d_2=\d_3=\cdots=\d_{p-q+1}$. The solution
(7) therefore takes the form, \bea d{s}^2 &=& F^{\frac{p+1}{8}}
(H{\tilde {H}})^{\frac{2}{7-p}} \left(\frac{H}{\tilde
H}\right)^{\frac{(p-q)\d_1}{8} + \frac{(p-q)(3- p)\d_2}{2 (7 -
p)}} \left(dr^2 + r^2 d\Omega_{8-p}^2\right)\nn & & +
F^{-\frac{7-p}{8}} \left(\frac{H}{\tilde
H}\right)^{\frac{(p-q)\d_1}{8} + \frac{(p-q)\d_2}{2}}\left(-dt^2 +
\sum_{i=1}^q (dx^i)^2 \right)\nn & & + F^{-\frac{7-p}{8}}
\sum_{j=q+1}^{p}\left(\frac{H}{\tilde H}\right)^{
\frac{\d_1}{8}(p-q - 8) + \frac{(p-q - 4)\d_2}{2}} (dx^{j})^2\nn
e^{2\phi} &=& F^{\frac{3-p}{2}} \left(\frac{H}{\tilde
H}\right)^{\frac{\d_1}{2}(4-p+q) - 2(p-q)\d_2}\nn F_{[8-p]} &=& b
{\rm Vol}(\Omega_{8-p}) \eea Note that we have removed `tilde'
from $d\tilde{s}^2$ and $e^{2\tilde \phi}$ for brevity. The
parameter relations (3) now take the forms, \bea & & \a-\b =
a\d_1\nn & &\frac{1}{2} \d_1^2 + \frac{1}{2} \a (\a-a\d_1) +
\frac{(p-q)(p-q-1)}{7-p}\d_2^2 =
\left(1-(p-q)\d_2^2\right)\frac{8-p}{7-p}\nn & & b = (7-p)
\omega^{7-p} (\a+\b) \sinh2\theta \eea where the functions $F$ and
$H$, $\tilde H$ are as given in eq.(2). It is clear from the above
that the metric in (8) has an isometry ISO($1,q$) $\times$
SO($p-q$) $\times$ SO($9-p$) and therefore the solution can be
interpreted as intersecting non-susy $p$-brane with non-susy
$q$-brane. It is clear that since we have $F_{[8-p]}$ which is in
general non-zero so, the non-susy $p$-brane is charged under it,
but since there is no $F_{[8-q]}$ in the solution, the charge of
the non-susy $q$-brane is zero. The solution is characterized by
four unknown independent parameters $\omega$, $\theta$, $\d_1$ and
$\d_2$. So, (8) represents the four parameter solution of
intersecting non-susy $p$-brane with chargeless non-susy
$q$-brane. Although we have concluded that (8) represents the
required solution from the symmetry of the metric, in order to
convince ourselves we will compare it with the intersecting brane
solutions obtained earlier for $p-q=4$ \cite{Bai:2005jr} since this is only case
where explicit solution is known. For this purpose we will define
another function $\hat F$ by, \be F = \left(\frac{H}{\tilde
H}\right)^\a \cosh^2\theta - \left(\frac{\tilde H}{H} \right)^\b
\sinh^2\theta\ = \hat F \,\left(\frac{H}{\tilde
H}\right)^{\tilde\a} \ee where \be \hat F = \left(\frac{H}{\tilde
H}\right)^{\a_1} \cosh^2\theta - \left(\frac{\tilde H}{H}
\right)^{\b_1} \sinh^2\theta\ \ee where we have defined $\a_1 +
\tilde \a = \a$ and $\b_1 - \tilde \a = \b$. Now it can be checked
that by defining the parameters as, \bea \tilde\a &=&
\frac{7-p}{7-q} \a_2 - \frac{p-q}{7-q}\d\nn \d_2 &=&
-\frac{7-p}{2(7-q)} (\a_2 + \d)\nn \d_1 &=& -\frac{p-q}{7-q} \a_2
+ \frac{7-p}{7-q}\d \eea we can recast the solution (8) as
follows, \bea ds^2 &=& \hat F^{\frac{p+1}{8}}\left(H\tilde
H\right)^{\frac{2}{7-p}} \left(\frac{H}{\tilde H}
\right)^{\frac{q+1}{8}\a_2}\left( dr^2 + r^2
d\Omega_{8-p}^2\right) \nn &+& \hat F^{-\frac{7-p}{8}}
\left(\frac{H}{\tilde H} \right)^{-\frac{7-q}{8}\a_2} \left(-dt^2
+ \sum_{i=1}^q (dx^i)^2 \right) + \hat F^{-\frac{7-p}{8}}
\left(\frac{H}{\tilde H} \right)^{\frac{q+1}{8}\a_2}
\sum_{j=q+1}^p (dx^j)^2 \nn e^{2\phi} &=& \hat
F^{\frac{3-p}{2}}\left(\frac{H}{\tilde
H}\right)^{\frac{3-q}{2}\a_2 +2\d}\nn F_{[8-p]} &=& b {\rm
Vol}(\Omega_{8-p}) \eea The parameter relations as obtained from
(9) now takes the form, \bea & & \a_1-\b_1 =
\left(\frac{p-q}{2}-2\right)\a_2 + \frac{p-3}{2} \d\nn & & b =
(7-p) \omega^{7-p} (\a_1 + \b_1) \sinh2\theta\nn & &
\frac{1}{2}\left(\d - \frac{p-3}{4} \a_1 -
\frac{q-3}{4}\a_2\right)^2\nn & & = \frac{8-p}{7-p} -
\frac{(p+1)(7-p)}{32}\a_1^2 - \frac{(q+1)(7-q)}{32}\a_2^2 -
\frac{(q+1)(7-p)}{16}\a_1\a_2 \eea The solution (13) is
characterized by four independent parameters $\omega$, $\theta$,
$\d$ and $\a_2$. Now comparing (13) with the intersecting non-susy
D$p$/D$(p-4)$ solution given in eq.(1) of ref.\cite{Bai:2005jr}, we find that
they exactly match for $q=p-4$ and when $\theta_1=\theta$,
$\theta_2=0$ indicating that the non-susy D$(p-4)$ brane is
chargeless. The parameter relation (14) also match with the
parameter relations given there. Thus comparing the explicit
solution of intersecting non-susy D$p$/D$(p-4)$, we conclude that
(8) indeed represents intersecting non-susy $p$-brane with
chargeless non-susy $q$-brane solution.

\section{Intersecting solution and black $p$-brane}

In this section we will show that when $q=0$, either of the
intersecting solution (8) or (13) can be mapped to the standard
black $p$-brane solution
\cite{Horowitz:1991cd,Duff:1993ye,Duff:1994an} by a coordinate
transformation for certain choice of two of the four parameters
characterizing the solution. For this purpose let us first write the
solution (13) for general $p$ and $q=0$ as follows, \bea ds^2 &=&
\hat F^{\frac{p+1}{8}}\left(H\tilde H\right)^{\frac{2}{7-p}}
\left(\frac{H}{\tilde H} \right)^{\frac{1}{8}\a_2}\left( dr^2 + r^2
d\Omega_{8-p}^2\right) \nn &-& \hat F^{-\frac{7-p}{8}}
\left(\frac{H}{\tilde H} \right)^{-\frac{7}{8}\a_2} dt^2 + \hat
F^{-\frac{7-p}{8}} \left(\frac{H}{\tilde H}
\right)^{\frac{1}{8}\a_2} \sum_{j=1}^p (dx^j)^2 \nn e^{2\phi} &=&
\hat F^{\frac{3-p}{2}}\left(\frac{H}{\tilde
H}\right)^{\frac{3}{2}\a_2 +2\d}\nn F_{[8-p]} &=& b {\rm
Vol}(\Omega_{8-p}) \eea where the parameters satisfy \bea & &
\a_1-\b_1 = \left(\frac{p}{2}-2\right)\a_2 + \frac{p-3}{2} \d\nn & &
b = (7-p) \omega^{7-p} (\a_1 + \b_1) \sinh2\theta\nn & &
\frac{1}{2}\left(\d - \frac{p-3}{4} \a_1 +
\frac{3}{4}\a_2\right)^2\nn & & = \frac{8-p}{7-p} -
\frac{(p+1)(7-p)}{32}\a_1^2 - \frac{7}{32}\a_2^2 -
\frac{7-p}{16}\a_1\a_2 \eea Let us now make a coordinate
transformation from $r$ to $\rho$ as follows, \be r = \rho
\left(\frac{1+\sqrt{f}}{2}\right)^{\frac{2}{7-p}} \ee where we have
defined, \be f = 1 - \frac{4\omega^{7-p}}{\rho^{7-p}} \equiv 1 -
\frac{\rho_0^{7-p}}{\rho^{7-p}} \ee From eq.(17) we find \bea H &=&
\frac{2}{1+\sqrt{f}}\nn \tilde H &=& \frac{2\sqrt{f}}{1+\sqrt{f}}
\eea where $H$ and $\tilde H$ are as given before in eq.(2). From
here we also obtain \be \frac{H}{\tilde H} = f^{-\frac{1}{2}} \ee
Now from the relation (11) we write \be \hat F = \left[\cosh^2\theta
- \sinh^2\theta \left(\frac{\tilde H}{H}\right)^{\a_1+\b_1}
\right]\left(\frac{H}{\tilde H}\right)^{\a_1} \ee When
$\a_1+\b_1=2$, the relation (21) reduces to \bea \hat F &=&
\left(\cosh^2\theta - \sinh^2\theta f\right) f^{-\frac{\a_1}{2}}\nn
&=& \left(1 + \frac{\rho_0^{7-p} \sinh^2\theta}{\rho^{7-p}}\right)
f^{-\frac{\a_1}{2}}\nn &=& {\bar H} f^{-\frac{\a_1}{2}} \eea where
we have defined ${\bar H} = 1 +
\sinh^2\theta\rho_0^{7-p}/\rho^{7-p}$. Also using \bea H \tilde H
&=& \frac{4\sqrt{f}}{1+f+ 2\sqrt{f}}\nn dr &=&
\frac{1}{\sqrt{f}}\left\{\frac{1+\sqrt{f}}{2}\right\}^{\frac{2}{7-p}}\,d\rho
\eea we find that the metric in (15) would reduce to the following
black $p$-brane form, \be ds^2 = {\bar
H}^{\frac{p+1}{8}}\left(\frac{1}{f} d\rho^2 + \rho^2 d\Omega_{8-p}^2
\right) + {\bar H}^{-\frac{7-p}{8}}\left(-f dt^2 + \sum_{j=1}^p
(dx^j)^2\right) \ee if the parameters $\a_1$ and $\a_2$ satisfy, \be
\a_1 = \frac{2}{7-p},\qquad {\rm and} \qquad \a_2 = 2. \ee Now from
the parameter relation (16) we also obtain the value of $\d$ as \be
\d = -\frac{2(6-p)}{7-p} \ee and the so the dilaton in (15) reduces
to \be e^{2\phi} = {\bar H}^{\frac{3-p}{2}} \ee The charge relation
in (16) becomes \be b = 2 (7-p) \omega^{7-p} \sinh2\theta =
\frac{1}{2} (7-p)\rho_0^{7-p} \sinh2\theta \ee The eqs.(24), (27)
along with (28) represent the known black $p$-brane solution
\cite{Horowitz:1991cd, Duff:1993ye,Duff:1994an}. Note that in
obtaining the black $p$-brane solution we have used the intersecting
solution (15) i.e. we have used (13) with general $p$ and $q=0$. The
parameters in this solution are $\omega$, $\theta$, $\d$ and $\a_2$
($\a_1$ and $\b_1$ are not independent). However, black $p$-brane
solution could also be obtained using intersecting solution (8)
where the unknown parameters are $\omega$, $\theta$, $\d_1$ and
$\d_2$ (here $\a$ and $\b$ are not independent). Using (12) in (25)
and (26) we can obtain the value of the parameters in solution (8)
such that it represent the black $p$-brane solution. The values are,
\bea \a + \b &=& 2,\qquad \a\,\,=\,\,\frac{16}{7}, \qquad
\b\,\,=\,\, -\frac{2}{7}\nn \d_1 &=& -\frac{12}{7},\qquad \d_2
\,\,=\,\, -\frac{1}{7} \eea Note that unlike in the previous case,
the values of the parameters here are universal in the sense that
they are independent of $p$ and will be useful for our discussion of
describing the two branches in the parameter space in the next
section.

Before closing this section let us make a few comments on the solution
in general.

\begin{itemize}

\item Note that the function $F$ or $\hat F$ as defined in eq.(2) or eq.(11) has
to be non-negative in general if the metric of the intersecting
brane solution has to remain real. Now for general non-vanishing
$\theta$ one can write $F = \left[\cosh^2\theta - \sinh^2\theta
\left(\frac{\tilde H}{H}\right)^{\a+\b}\right]
\left(\frac{H}{\tilde H}\right)^{\a}$ and similarly for $\hat F$.
This implies that $\a+\b \geq 0$ (or similarly $\a_1+\b_1 \geq
0$), otherwise $F$ can not be kept non-negative once $r$
approaches $\omega$. So, we need in general $\a+\b \geq 0$, but
for $\theta=0$, we can  have $\a+\b <0$ as well.

\item Given our understanding that the black D$p$-brane can be obtained from
charged or chargeless non-susy D$p$-brane with intersecting chargeless non-susy
D0-brane only when $\a+\b=2>0$ and $\d_1=-12/7$, $\d_2=-1/7$, independent of the
choice of
other parameters, implies that black brane occurs only in the $\a+\b\geq 0$-branch
even though the other branch is allowed when $\theta=0$.

\item  Let us now focus on $\a+\b \geq 0$-branch. Even for this branch the black
D$p$-brane occurs only for
particular choice of the other parameters $\a=16/7$, $\d_1=-12/7$ and $\d_2=-1/7$.
Then the natural question is what the other values of the parameters correspond to?
Even for a given $\a+\b \geq 0$, we can still have two branches of values for other
parameters such as $\d_1$.

\item  In the $\a+\b \geq 0$-branch the black D$p$-brane has well-defined horizon
but for other values of the parameters we have null or naked
singularity for each configuration. One would expect that those with
these singularities should be more unstable than the one with
regular horizon because an observer at infinity will see the rapid
annihilation process of the brane-antibrane for the system without a
regular horizon than the one with a regular horizon. In the
supergravity description the appearance of singularities in either
case is due to supergravity approximation and in the full
non-perturbative string language, we do not expect the singularities
to form. The supergravity configurations only provide us an
indication which process we should assign to.

\item So, the above indicates that the parameter space of the (intersecting) non-susy brane
configuration has various disjoint branches, which may imply a phase
structure of the underlying dynamics. One branch gives rise to the
brane-antibrane annihilation where the underlying dynamics is
governed by the open string tachyon condensation \cite{Sen:2004nf}.
Under this process the brane-antibrane configuration evolves into
the final BPS brane or Minkowski vacuum depending on the initial
charge of the system \cite{Lu:2004dp,Bai:2005jr}. In this branch we
really do not have the condition of regular horizon formation to
begin with except for the special complete chargeless case as
mentioned earlier and to be discussed in the next section, and the
solutions will have null or naked singularity (although this should
be understood as the artifact of perturbative string theory). The
other branch can give rise to the regular horizon formation and then
the underlying dynamics is governed by possible closed string
tachyon condensation \cite{Horowitz:2005vp,Horowitz:2006mr}. Under
this process the black $p$-brane would evolve into bubble (so no
singularity) according to Horowitz's prescription.

\item Let us recall that the static non-susy $p$-brane or
D$p$/${\bar {\rm D}}p$ brane configuration obtained in
ref.\cite{Lu:2004ms}, containing a parameter $\delta$ which was
bounded from above for the solution to remain real. Right at the
bounded value, the corresponding configuration has a null-singular
horizon. We also noted that when the parameter exceeds the bound the
solution becomes complex in general (except for a particular choice
of $\theta$ where the configuration has periodically naked
singularities), therefore, signalling a possible phase transition.
This feature persists even for the intersecting non-susy brane
solutions. In the next section we will demonstrate the similar
characteristic, indicating the existence of two possible phases, for
the two examples, namely for D0/${\bar {\rm D}}$0 and intersecting
D4/${\bar {\rm D}}$4 with chargeless D0/${\bar {\rm D}}$0 system
where the explicit microscopic descriptions are known.

\end{itemize}

\section{Dynamical structure of non-susy D$p$-branes}

In this section we give some evidence that the non-susy D$p$-branes
have two disjoint dynamical structures, one corresponding to the
brane-antibrane annihilation process and the other related to the
black $p$-brane dynamics. To illustrate the points we use the
microscopic representation of the SUGRA parameters in terms of the
physical parameters of the non-susy D$p$-branes. These are known
only for the D0/${\bar {\rm D}}$0 system and the intersecting
D4/${\bar {\rm D}}$4 with chargeless D0/${\bar {\rm D}}$0 system and
we discuss the two cases separately below.

\vspace{.5cm}

\noindent {\it (a) {\rm D0}/${\bar {\rm D}}{\rm 0}$ system}

\vspace{.2cm}

The supergravity solution of D0/${\bar {\rm D}}$0 system (when it is
chargeless it can represent either the equal number of D0/${\bar
{\rm D}}$0 system or non-BPS D0-branes) can be obtained from eq.(8)
by putting $p = q = 0$. This solution as discussed in
ref.\cite{Lu:2004ms} is characterized by three independent SUGRA
parameters $\omega$, $\theta$ and $\delta_1$. In
ref.\cite{Lu:2004dp} we have related these parameters to the
microscopic physical parameters namely the number of D0-branes
($N$), the number of ${\bar {\rm D}}$0-branes (${\bar N}$) and
the tachyon parameter ($T$) of the system. When we related the
supergravity solution this way we assumed that the solution (8)
represents one of a continuous family of the brane-antibrane
configurations labeled by the tachyon parameter ($T$) in the
brane-antibrane annihilation process. We also noted that the
parameter $\delta_1$ has to be bounded in order for this to be true.
We noticed that when $\d_1$ exceeds that bound, the solution becomes
complex and thereby signalling a possible phase transition in the
system. In order for describing the brane-atibrane annihilation
process through tachyon condensation such that the system evolves
either to a BPS D$p$-brane or Minkowski vacuum we found that the
parameter $\d_1$ must be given as \cite{Lu:2004dp}, \bea
\delta_1^{(-)} &= &\frac{1}{2 c_p}
\frac{a}{|a|}\left\{|a|\sqrt{\cos^2 T +\frac{(N - \bar N)^2}{4
N{\bar N} \cos^2 T}}\right.\nn &\,& \left. - \sqrt{a^2 \left(\cos^2
T +\frac{(N - \bar N)^2}{4 N{\bar N} \cos^2 T}\right) + 4
\left(\frac{2(8 - p)}{7 - p} c^2_p - \cos^2 T\right)}\right\} \eea
where $a=(p-3)/2$ and $c_p$ is an unknown constant depending on $p$
but is bounded as \be c_p \geq  \frac{7 - p}{4} \sqrt{\frac{p +
1}{2(8 - p)}} \ee in order for $\delta_1^{(-)}$ to be real. It is
not difficult to check that the absolute value of $\d_1^{(-)}$,
i.e.,  $|\delta_1^{(-)}|$, in (30) is bounded by,
\be |\delta_1^{(-)}| \leq \frac{4}{7 -
p}\sqrt{\frac{2(8 - p)}{p + 1}} \ee for which the configuration
remains real, and approaches zero (when $N \neq \bar N$ and $p \neq
3$) or the maximum value $\sqrt{2(8 - p)/(7 - p)}$ less than the
bounded one given in eq.(32) (when $N = \bar N$ or $p = 3$) as $T
\to \pi/2$. Note that in ref.\cite{Lu:2004dp}, for simplicity, $c_p$
was taken as $c=\sqrt{(7-p)/2(8-p)}$. Now since $\d_1$ was
obtained from a quadratic relation (9) (with $q=0$) it actually has
two roots and we just kept the above bounded root there. This
corresponds to one branch value of $\d_1$ and the other branch value
of $\d_1$ is given by the other root as \bea \delta_1^{(+)}
&=&\frac{1}{2 c_p} \frac{a}{|a|}\left\{|a|\sqrt{\cos^2 T +\frac{(N -
\bar N)^2}{4 N{\bar N} \cos^2 T}}\right. \nn &\,&\left. + \sqrt{a^2
\left(\cos^2 T +\frac{(N - \bar N)^2}{4 N{\bar N} \cos^2 T}\right) +
4 \left(\frac{2(8 - p)}{7 - p} c^2_p - \cos ^2 T\right)}\right\}
\eea which, on the other hand, exceeds the bounded value in general,
and for this reason it was dropped in ref.\cite{Lu:2004dp}. We know
that the first expression gives a good description of the
brane-antibrane annihilation leading to either a BPS D$p$-brane or
Minkowski vacuum i.e. the open string tachyon condensation. From our
previous discussion the second expression will be responsible for
the unbounded $\d_1$ and therefore may be related to a possible
phase transition. Further, as will be shown, $\d_1 = - 12/7$ for
the usual black 0-brane is a solution of $\d_1^{(+)}$ but not
$\d_1^{(-)}$ in general except for a degenerate case\footnote{As
will be clear from the following discussion, the two branches are
disjoint when $N \neq \bar N$, otherwise they will meet at a point
at $T = 0$, i.e., at the top of the tachyon potential as claimed in
the introduction. One can check this explicitly by noting that in
order to have $\d_1^{(-)} = \d_1^{(+)}$, the second square root in
each expression must vanish which can occur only as $N = \bar N$ and
$c_p$ takes its minimal value given in (31).}. This seems also
consistent with the possible phase transition for black branes
through the closed string tachyon condensation to bubbles of nothing
as proposed by Horowitz and
others\cite{Horowitz:2005vp,Horowitz:2006mr}. Since we propose black
brane to appear in some branch of the parameter space and since only
for $p=0$, the non-susy $p$-brane solution (which is obtained from
(8) by putting $q=0$ as well) can be mapped to black D0-brane so we
will use $p=0$ in (30) and (33). Now we will examine explicitly
which of the above $\d_1$ eq.(30) or (33) will satisfy the condition
for the black 0-brane i.e. $\d_1=-12/7$ (see eq. (29)) for certain
value of the tachyon. To discuss this let us first set $N=\bar N$,
i.e., the chargeless case ($N \neq \bar N$ will be discussed later).
Since $a=-3/2$ for 0-brane, putting $\d_1^{(-)}=-12/7$ in (30) we
get \be \frac{24 c_0}{7} - \frac{3}{2}\cos T = -
\sqrt{\frac{9}{4}\cos^2 T + 4\left(\frac{16}{7} c_0^2 - \cos^2
T\right)} \ee Now from eq.(31) we have $c_0\geq 7/16$, the lhs of
eq.(34) gives $24c_0/7 - 3\cos T/2 \geq (3/2)(1-\cos T) \geq 0$. So,
in general this equation can not be satisfied. The only way it can
be satisfied is if the rhs of (34) vanishes. It can be easily
checked that for this to happen $c_0$ must take the minimum value
$7/16$ and $\cos T=1$, i.e. the tachyon is at the top of the
potential, a sort of critical value. For this $c_0$, we have from
(30) \be \d_1^{(-)} = -\frac{4}{7}\left(3\cos T - \sqrt {7} \sin
T\right) \ee We thus see from this expression that $\d_1^{(-)} =
-12/7$ is the smallest value and any other value of $\d_1^{(-)}$ in
the condensation will be larger. The implication of this is that
just as the brane-antibrane starts to annihilate each other, the
system has a choice to either form a regular horizon (then the
brane-antibrane can no longer annihilate each other) or to
annihilate each other without forming a regular horizon. In the
former case an analog of a transition to ``bubble of
nothing"\footnote{We will address the possible transition for
0-brane case in a forthcoming paper as will be mentioned later in
this section.} may take over while for the latter it is just the open
string tachyon condensation process. For all other values of $c_0$,
we have only brane-antibrane annihilation via open string tachyon
condensation and there is no regular horizon formation in this
branch since $\d_1^{(-)}=-12/7$ can never be satisfied.

Now we consider the second expression (33) for the black brane and
in this case we get for $\d_1^{(+)}=-12/7$, \be \frac{24 c_0}{7} -
\frac{3}{2}\cos T =  \sqrt{\frac{9}{4}\cos^2 T + 4\left(\frac{16}{7}
c_0^2 - \cos^2 T\right)}. \ee This equation can be solved with
either $\cos T = 16 c_0/7$ or $\cos T = 2c_0/7$. The first solution
can only be satisfied if $c_0=7/16$ (since this is the minimum value
of $c_0$) and $\cos T = 1$ which is the critical case as mentioned
above but there is some complication here. Let us discuss this in a
bit detail. For $c_0 = 7/16$ we have from (33) \be \d_1^{(+)} = -
\frac{4}{7}\left(3\cos T + \sqrt{7} \sin T\right).\ee It is not
difficult to check that this $\d_1^{(+)}$ has the smallest value of
$- 16/7$ at $\cos T = 3/4$, i.e. reaching its bounded value, and the
largest value of $- 4/\sqrt{7}$ at $\cos T = 0$ or $T = \pi/2$.
Since $- 4/\sqrt{7} > - 12/7$, we therefore expect another solution
for $\delta^{(+)}_1 = - 12/7$ between $\cos T = 3/4$ and $\cos T =
0$ in addition to the $\cos T = 1$. Indeed one can find that this
occurs at $\cos T = 1/8$ from the above equation which is just given
by $\cos T = 2 c_0/7$ for $c_0 = 7/16$. Given that $|\d_1^{(+)}|
\leq 16/7$, i.e., its bounded value, for $ 0 \leq T \leq \pi/2$, we
therefore expect that this $\d_1^{(+)}$ for the chargeless case can
also be used to describe the brane-antibrane annihilation or the
open string tachyon condensation. The actual process may behave like
this: just as the brane-antibrane starts to annihilate each other,
the system has a choice  either to form a regular horizon (then the
brane-antibrane can no longer annihilate each other and the possible
phase transition associated with the black brane is taken over) or
to annihilate each other without forming a regular horizon. When the
system makes a choice of annihilating each other initially, i.e., at
$\cos T = 1$, it will continue so until $\cos T = 1/8$. At this
point, the system has again a choice either to form a regular
horizon without continuing the annihilation or to continue the
annihilation to end up with a Minkowski vacuum. We now consider
$7/16 < c_0 \leq 7/2$. The only solution to (36) is $\cos T = 2
c_0/7$ as given above. We now have from (33) \be \d_1^{(+)} = -
\frac{1}{2 c_0}\left( \frac{3}{2} \cos T + \sqrt{\frac{64}{7} c_0^2
- \frac{7}{4} \cos^2 T}\right).\ee The $\d_1^{(+)}$ has a minimum
value of $- 16/7$, which is also its bounded value, at $\cos T = 12
c_0/7$ if $7/16 < c_0 \leq 7/12$. Given that there is only one
solution $\cos T = 2 c_0/7$ now and $|\d_1^{(+)}| \leq 16/7$, we
therefore expect the following process: the brane-antibrane system
initially annihilates each other until $\cos T = 2 c_0/7$. At this
point, it has a choice either to form a regular horizon without
annihilating each other further or to continue to annihilate each
other without forming a regular horizon to end up with a Minkowski
vacuum.

Let us next discuss the case when $N\neq \bar N$. For this case the
condition for the black brane $\d_1=-12/7$ gives from the first
expression (30), \be \frac{24 c_0}{7} = \frac{3}{2} \sqrt{\cos^2 T
+\frac{(N - \bar N)^2}{4 N{\bar N} \cos^2 T}} -
\sqrt{\frac{9}{4}\left(\cos^2 T +\frac{(N - \bar N)^2}{4 N{\bar N}
\cos^2 T}\right) + 4 \left(\frac{16}{7} c_0^2 - \cos^2 T\right)}.
\ee We have to see whether this equation can be made consistent for
the allowed values of $c_0$ and some values of $\cos T$. In order
for eq.(39) to be consistent, we need to have the second term on
the right smaller than the first term and this can only be true if
the second term in the square root is negative, i.e., \be
\frac{16}{7} c^2_0 - \cos^2 T < 0 \ee or $ 0 < c_0 < \sqrt{7} \cos T
/4$. But actually for consistency we have to have the following
equation to be satisfied (which can be derived from (39)) \be
\frac{(N - \bar N)^2}{4 N{\bar N} \cos^2 T} = \frac{(\frac{16}{7}c_0
- \cos T)(\frac{2}{7} c_0 - \cos T)(\frac{16}{7}c_0 + \cos
T)(\frac{2}{7} c_0 + \cos T)}{(\frac{18}{7} c_0)^2} \ee The left
side of the above equation is positive and so must be the right
side. Given that $c_0 < (\sqrt{7}/4)\cos T$, i.e., $2c_0 /7 <
(1/2\sqrt{7}) \cos T < \cos T$, we must have from (41) \be
\frac{16}{7} c_0 - \cos T < 0 \ee But the condition (42) gives $c_0
< (7/16) \cos T $ which is even more constrained than the bound $c_0
< (\sqrt{7}/4)\cos T$. In other words, for consistency we must have
this constraint to be satisfied but this one contradicts with the
previous constraint $c_0 \geq 7/16$ derived earlier. Therefore we
can not make (39) consistent implying that there is no solution from
the expression (30) for $\delta_1^{(-)}$ to form a regular horizon
which is the expected result. For the second expression (33) for
$\delta_1^{(+)}$, we expect in general a solution since the relevant
equation is \be \frac{24 c_0}{7} = \frac{3}{2} \sqrt{\cos^2 T
+\frac{(N - \bar N)^2}{4 N{\bar N} \cos^2 T}} +
\sqrt{\frac{9}{4}\left(\cos^2 T +\frac{(N - \bar N)^2}{4 N{\bar N}
\cos^2 T}\right) + 4 \left(\frac{16}{7} c_0^2 - \cos^2 T\right)} \ee
Now we still have $c_0 \geq 7/16$ but here we don't need to have
$c_0 < (\sqrt{7}/4)\cos T$ as before for consistency. But we still
have the equation (41) to be satisfied. So the condition for the
right side to be positive and consistent with $c_0 \geq 7/16$ is
$c_0 > 7/2$. Since the term in the square root on the right side of
(43) is $(64/7) c_0^2$ whose square root is less than the left side
of this equation, therefore we are certain that there must be at
least a solution of this equation for $\cos T$ as expected. Moreover
from (33) for $p = 0$, $|\d_1^{(+)}|$ for the charged case will
be unbounded for small $\cos T$ and therefore we expect that this
branch of $\d_1^{(+)}$ will be responsible for possible phase
transition related to the black brane such as from black brane to
``bubble of nothing".

\vspace{.5cm}

\noindent {\it (b) Intersecting {\rm D4}/${\bar {\rm D}}{\rm 4}$ with chargeless
{\rm D0}/${\bar {\rm D}}{\rm 0}$ system}

\vspace{.2cm}

The supergravity solutions of intersecting D$p$/${\bar {\rm D}}p$
with D$(p-4)$/${\bar {\rm D}}(p-4)$ system
where both brane-antibrane systems are charged are given in ref.\cite{Bai:2005jr}.
These
solutions are characterized by five parameters $\omega$, $\theta_1$, $\theta_2$,
$\Delta_1(=\a_1+\b_1)$ and $\Delta_2(=\a_2+\b_2)$. We would like to refer the reader
to \cite{Bai:2005jr} for
the details of the solution, but we just mention that the subscript 1 and 2 refer to
D$p$/${\bar {\rm D}}p$ system and D$(p-4)$/${\bar {\rm D}}(p-4)$ system respectively.
These supergravity parameters
were related in \cite{Bai:2005jr} to physical microscopic parameters,
namely, the  number
of D$p$-branes ($N_1$),
the number of ${\bar {\rm D}}p$-branes ($\bar{N}_1$), the number of D$(p-4)$-branes
($N_2$), the number of ${\bar {\rm D}}(p-4)$-branes ($\bar{N}_2$) and the tachyon
parameter
($T$) such that we get the correct picture of brane-antibrane annihilation
via tachyon condensation. These relations are not known for other intersecting brane
solutions. Since the black-brane appears when one of the intersecting branes is
the chargeless 0-brane/anti-0-brane, we will make use of the solutions of
ref.\cite{Bai:2005jr}
only for $p=4$ after discussing some generalities. From our experience of tachyon
condensation on the simple D$p$/${\bar {\rm D}}p$ system and the fact that there is
no interaction between the D$p$/${\bar {\rm D}}p$ and D$(p-4)$/${\bar {\rm D}}(p-4)$
system we found in (see eq.(10) of \cite{Bai:2005jr}) that the parameters $\Delta_1$ and
$\Delta_2$ must satisfy,
 \bea
& &\Delta_1 \sqrt{1 + \frac{(N_1 - \bar
 N_1)^2}{c_p^2 A^2 \Delta_1^2 \cos^2 T}} \,\, =\,\,
 \pm\left[\frac{1}{c_p A} \sqrt{\frac{(N_1 -\bar N_1)^2}{\cos^2 T} + 4 N_1
 \bar N_1 \cos^2 T} - \frac{p-3}{2}\delta\right]\nn
& &\Delta_2 \sqrt{1 + \frac{a^2 (N_2 - \bar
 N_2)^2}{c_p^2 A^2 \Delta_2^2 \cos^2 T}} \,\,=\,\,
\pm \left[\frac{a}{c_p A} \sqrt{\frac{(N_2 -\bar N_2)^2}{\cos^2 T} + 4 N_2
 \bar N_2 \cos^2 T} - \frac{p-7}{2}\delta\right]
 \eea
where in the above $A=\sqrt{N_1\bar{N}_1} + a \sqrt{N_2\bar{N}_2}$ and
$a=(2\pi\sqrt{\a'})^4/V_4$ with $V_4$ the volume of the compact directions.
Now using the parameter relation analogous to (9) or (14) in that case
given by,
\be
\Delta_1^2 + \Delta_2^2 + \left(4-\frac{(p-3)^2}{4}-\frac{(p-7)^2}{4}\right)
\d^2 = 8 \frac{8-p}{7-p}
\ee
we obtain a quadratic relation in $\d$, after substituting $\Delta_1$ and $\Delta_2$
from (44) in (45). So, the equation has two roots given by,
\bea
& &\delta^{(-)} \,=\, \frac{1}{8 c_p A} \frac{K}{|K|}
\left[|K| - \sqrt{K^2 + 64
 \left(\frac{2(8 - p)}{7 - p} c_p^2 A^2- \left(N_1\bar N_1 + a^2 N_2 \bar N_2\right)
\cos^2 T\right)}\right]\nn
& & \delta^{(+)} \,=\, \frac{1}{8 c_p A} \frac{K}{|K|} \left[|K| + \sqrt{K^2 + 64
 \left(\frac{2(8 - p)}{7 - p} c_p^2 A^2- \left(N_1\bar N_1 + a^2 N_2 \bar N_2\right)
\cos^2 T\right)}\right]\nn
\eea
where,
\be K =  (p - 3) \sqrt{\frac{(N_1 - \bar N_1)^2}{\cos^2 T} + 4 N_1\bar N_1 \cos^2 T}
 + a (p - 7)\sqrt{\frac{(N_2 - \bar N_2)^2}{\cos^2 T} + 4 N_2\bar N_2 \cos^2 T}
\ee In (46) $\d^{(-)}$ is bounded while $\d^{(+)}$ is not when
$\cos T$ approaches zero and so only the first one is responsible
for the open string tachyon condensation when either $N_1 \neq
\bar N_1$ or $N_2 \neq \bar N_2$ or both, i.e., the charged case.
Let us point out some features of the solutions (46). First, the
term in the square root should be positive for reality of the
solutions. This in general sets the bound for $c_p$. Second, in
order to get black D$p$-brane, we know from (26) that $\d = - 2(6
- p)/(7 - p) <0$ and this implies that we need $K<0$ for at least
some values of $\cos T$. It is obvious that when $K>0$, the second
equation in (46) can not be satisfied for the negative $\d$, but
this is not obvious from the first one. We will show that even for
this case $\d$ can not be negative, therefore no solution exists
for this $\d$ when $K > 0$.

Let us now put $p=4$ and for our purpose of making connection with
black D4-brane we put $N_2=\bar{N}_2$. Let us also for simplicity
consider first the case of $N_1=\bar{N}_1$ i.e. the chargeless
D4/${\bar {\rm D}}$4 system. Unlike the D0-case discussed
previously, here we need to fix both $\Delta_1(=\a_1+\b_1)$ and
$\Delta_2(=\a_2+\b_2)$ to be 2 and also from (26) $\d=-4/3$.
Indeed we note from the general system given in (15) and (16) that
we have $\Delta_1 = \a_1+\b_1=2$ and $\Delta_2 \equiv \a_2 =2$ and
from (26) $\d=-4/3$. Now by using the first equation of (44) for
the present case, we get \be \frac{N_1}{c_4A}\cos T = \frac{2}{3}
\ee Note that we have used the plus sign of the rhs of the
equation since with minus sign this equation can not be satisfied.
On the other hand from the second equation of (44) we get, \be
\frac{aN_2}{c_4A}\cos T = 2 \ee Here also plus sign is chosen as
with the minus sign the equation can not be satisfied except for
some degenerate case which we ignore. These two imply that the
solution exists only if $aN_2=3N_1$ and now $\cos T = 8c_4/3$.
Note that we have $A=4N_1$ and $K=2(N_1-3aN_2)\cos T = -16N_1 \cos
T < 0$ and this is consistent with our above observation. Now in
order for $\d$'s to remain real (i.e. for the quantity inside the
square root of eq.(46) to remain positive) we have from (46) \be
c_4^2 \geq \frac{3}{8
 A^2}\left[\left(N_1^2 + a^2 N_2^2\right) - \frac{(N_1 - 3 a
 N_2)^2}{16}\right]\cos^2 T
 \ee
Since $c_4$ is a constant, independent of the $N$'s and the value of $\cos T$,
we need to have the following condition so that (50) can always be satisfied,
\be
c_4^2 \geq \frac{3}{8
 A^2}\left[\left(N_1^2 + a^2 N_2^2\right) - \frac{(N_1 - 3 a
 N_2)^2}{16}\right]
\ee Now let us see what is the value of $c_4$ for the present
solution where $aN_2=3N_1$. We find from (51) $c_4 \geq 3/8$. Now we
have to see whether this condition is consistent with the solution
namely, $\cos T = 8c_4/3$. It is obvious that this condition is
consistent only for the critical value $c_4 = 3/8$ and $\cos T =1$.
Actually for this case the square root in (46) vanishes and so
solves both the equations\footnote{Again we see a meet point at
$\cos T = 1$ of the two branches for the chargeless case.}. The
explanation for the underlying picture is the same as for the
D0-case i.e. the system starts at the top of the potential and it
has the choice of either forming the brane-antibranes and
subsequently annihilate each other through open string tachyon
condensation to end up with a Minkowski vacuum or forming the
regular horizon for the black D4 phase and subsequently decay to
bubbles through possible closed string tachyon condensation.

Let us next discuss the general case where $N_1 \neq \bar{N}_1$.
It is in general quite involved to discuss the solution directly
from (46) and much easier to use eq.(44) with $p=4$ instead. The
second equation in this case has exactly the same form as before
\be aN_2 \cos T(aN_2 \cos T - 2c_4A) = 0\ee where we have used
$\Delta_2 = 2$, and $\d=-4/3$. Note that $\cos T \neq 0$ comparing
the same $\cos T$ obtained from the first equation in (44). Now
since $(p-3)\delta/2 = -2/3 <0$, only the plus sign can be taken
in the first equation of (44) and the equation is, \be \sqrt{4 +
\frac{(N_1 - \bar N_1)^2}{c_4^2
 A^2 \cos^2 T}} - \frac{2}{3} = \sqrt{4 \frac{ 4 N_1\bar N_1}{a^2
 N_2^2} + \frac{(N_1 - \bar N_1)^2}{c_4^2 A^2 \cos^2 T}}
 \ee
where we have used $aN_2 \cos T = 2 c_4 A$ from eq.(52) in some of the terms above.
It is clear from (53) that in order to have a consistent equation
we must have the following constraint,
\be
\frac{4N_1\bar{N}_1}{a^2N_2^2} < 1
\ee
Further from the same equation (53) we also have,
\be
\frac{a^2 N_2^2 (N_1 -
 \bar N_1)^2}{16 (c_4^2 A^2)^2} = \left(\frac{2}{3} - \frac{6
 N_1\bar N_1}{a^2 N_2^2}\right)\left(\frac{8}{3} - \frac{6
 N_1\bar N_1}{a^2 N_2^2}\right)
\ee where the positivity of the lhs of the above equation gives
further, by considering (54), \be \frac{4N_1\bar{N}_1}{a^2N_2^2} <
\frac{4}{9} \ee and also we solve from  (53) the
following\footnote{Note that the brane number $N$'s and the
parameter $a$ appearing in (57) should be constrained such that
$\cos^2 T \leq 1$ and also the constraint for $c_4$ is not
violated.}, \bea \cos^2 T &=& \frac{|N_1 - \bar N_1|}{
 2 a N_2 \sqrt{(1/3 - 3 N_1\bar N_1/a^2 N_2^2)(4/3 - 3 N_1\bar N_1/a^2
 N_2^2)}},\nn
 c_4^2 A^2 &=& \frac{a N_2 |N_1 - \bar N_1|}{
 8 \sqrt{(1/3 - 3 N_1\bar N_1/a^2 N_2^2)( 4/3 - 3 N_1\bar N_1/a^2
 N_2^2)}},
\eea where we have also used $\cos T = 2 c_4 A/a N_2$ from eq.
(52). In obtaining (57), we have squared
 some equations and have not checked the consistencies
 that the term inside the square root in (46) remains
positive. We need to check a few things before we can be
 claim that the above solution indeed exists.  Let us discuss
under what conditions, a solution with $\delta < 0$ can exist.
Let us first check the value of $K$ in (47). In this case we have
 \be
K =   \sqrt{\frac{(N_1 - \bar N_1)^2}{\cos^2 T} + 4 N_1\bar N_1 \cos^2 T}
 - 6 a N_2 \cos T
\ee We see that $K$ is positively infinite
 when $\cos T \to 0$. So if the minimum value of $K$ is positive,
 then $K$ is always positive. If this happens, then we expect a
 possible solution from the first equation in (46) only when
 \be
\frac{8}{3} c_4^2 A^2 \geq \left(N_1\bar N_1 + a^2
 N_2^2\right) \cos^2 T
\ee When $K < 0$ for certain values of $\cos T$, then a solution
is possible from the first equation when the above inequality is
 reversed and from the second equation of (46) in general. So we
 now check the sign for $K$ from the solution (57). Using this
in (47) we get
\be
 K = - \frac{\sqrt{2 a N_2 |N_1 - \bar N_1|}}{\left[(1/3 - 3 N_1\bar N_1/a^2 N_2^2)
 ( 4/3 - 3 N_1\bar N_1/a^2 N_2^2)\right]^{1/4}} \left(\frac{7}{3}
 + \frac{3 N_1 \bar N_1}{a^2 N_2^2}\right) < 0
\ee
This shows that (57) may be a solution.
Before we make sure this and
 know which one in (46) actually solves this, let us evaluate the
 term with the square root in (46) and we find
 \bea
&& \sqrt{K^2 + 64 \left[\frac{8}{3} c_4^2 A^2 - \left(N_1\bar N_1 + a^2
 N_2^2\right)\cos^2 T\right]}\nn
 && = \frac{\sqrt{2 a N_2 |N_1 - \bar N_1|}}{\left[(1/3 - 3 N_1\bar N_1/a^2 N_2^2)
 ( 4/3 - 3 N_1\bar N_1/a^2 N_2^2)\right]^{1/4}} \left(\frac{1}{3}
 - \frac{3 N_1\bar N_1}{a^2 N_2^2}\right)
\eea where we made use of (56). From (60) and (61) one can check
 directly that the solution in (57) is indeed a solution of the
 second equation in (46) but not the first one. This is entirely
 consistent with our expectation since now $\cos T \neq 1$ and only the unbounded
channel (or the second solution of (46)) is associated with the
regular horizon formation as we conjectured.

As a closing remark in this section we emphasize that in this paper
we have made use of the semi-empirical relations, obtained
previously in refs.\cite{Lu:2004dp,Bai:2005jr}, between the SUGRA
parameters and microscopic physical parameters (number of branes,
number of antibranes and the open string tachyon parameter) of the
(intersecting) non-susy brane system to discuss the brane-antibrane
annihilation process and the black brane formation and the related
phase transition. In particular, for the charged system we have
shown that the specific value of the parameter $\d$ for the black
brane can only occur in the unbounded channel. Given that the
non-susy brane configuration represents the underlying
brane-antibrane system and in general $\cos T \neq 1$ for the black
brane value of $\d$ in the second expression, we can expect that the
brane-antibrane in this case starts to annihilate from $\cos T = 1$
to the value of $\cos T$ corresponding to the black brane value of
$\d$. At this particular value of $\cos T$, the system has a choice
of either forming a regular horizon (the possible closed string
tachyon condensation will be taken over as discussed by Horowitz to
end up with a bubble of nothing) or continuing the annihilation so
that $\d$ reaches its bounded value. Then also a possible phase
transition exists since the configuration becomes complex there once
$\d$ exceeds its bounded value and this is corroborated by the fact
that right at the bounded value of $\d$, the corresponding
configuration also has a horizon which is null-singular. We are
currently investigating the possible phase transitions related to
the different dynamics considered here. However, Horowitz's picture
of closed string tachyon condensation of black brane doesn't seem to
apply for the $p = 0$ case and and it appears that we need to lift
the configuration to eleven dimensions. Whether our current
investigation for $p = 0$ case can be of any help in understanding
the various phase transitions we mentioned is to be seen.

\section{Conclusion}

To conclude, contrary to the common belief, we have argued in this
paper that the non-susy (non-extremal) $p$-branes are of only one
type and not of two types. Usually it is known that there are two
possible ways of constructing the non-susy $p$-branes in type II
supergravities and they are referred to as type 1 and type 2 in
the literature \cite{Lu:1996er}. The type 1 metric has the form \be ds^2 =
e^{2A}\left(dr^2 + r^2 d\Omega_{8-p}^2\right) + e^{2B}\left(-dt^2
+ \sum_{i=1}^p (dx^i)^2\right) \ee in ten dimensions, where the
functions $A$ and $B$ are quite general but depends on $r$ only.
For the type 1 solution $A$, $B$ satisfy \be (p+1) B + (7-p) A =
\ln G \ee where $G(r)$ is an arbitrary function. Note that when
$G(r) = 1$, the type 1 solution reduces to the usual BPS $p$-brane
solution. In ref.\cite{Lu:2004ms} this solution is referred to as non-susy
$p$-brane solution and has been interpreted as
$p$-brane/anti-$p$-brane (or non-BPS $p$-brane) solution. On the
other hand the type 2 metric has the form, \be ds^2 =
e^{2A}\left(f^{-1}\, dr^2 + r^2 d\Omega_{8-p}^2\right) +
e^{2B}\left(- f\,dt^2 + \sum_{i=1}^p (dx^i)^2\right) \ee and now
the function $A$ and $B$ satisfy \be (p+1) B + (7-p) A = 0 \ee
just like a BPS $p$-brane. We have shown that type 2 solutions are
in fact contained in the generalized type 1 as a special case and
therefore we only have one type of solutions, namely, the type 1
solutions.

To show this we know that non-susy branes of type 1 can be of
different kinds. That is, there can be only one kind of brane
involved or there can be intersections of many kinds of branes,
where the different branes can be either charged or chargeless under
various form-fields. We have argued that the type 2 $p$-brane
solution can be regarded as intersecting non-susy $p$-brane with
chargeless non-susy 0-brane. We have shown this explicitly when the
type 2 solution is the usual black $p$-brane. So, we first
constructed an intersecting solution of non-susy $p$-brane with
chargeless non-susy $q$-brane i.e. of type 1. This is constructed
from a non-susy $q$-brane solution delocalized in $(p-q)$ transverse
directions and then applying T-dualities successively in all the
delocalized directions. The resulting solution has $(p-q+3)$
independent parameters and is anisotropic in the delocalized
directions. When we made the solution isotropic (in the delocalized
directions) by setting the parameters associated with these
directions equal, the solution has an isometry group ISO($1,q$)
$\times$ SO($p-q$) $\times$ SO($9-p$) and is dependent on four
independent parameters. From this isometry we recognized the
solution to be intersecting non-susy $p$-brane with chargeless
non-susy $q$-brane. The non-susy $q$-brane is chargeless because
there is no gauge field associated with it, unlike the non-susy
$p$-brane. The intersecting solution when both the branes are
charged is known explicitly only when $p-q=4$ or 0. We have compared
these solutions with the ones we obtained in this paper. Next we
showed that when we set $q=0$ i.e. when the isometry group becomes R
$\times$ ISO($p$) $\times$ SO($9-p$), the resulting solution can be
mapped to the standard black $p$-brane solutions by a coordinate
transformation when two of the four parameters ($\omega$, $\theta$,
$\d$, $\a_2$ or $\omega$, $\theta$, $\d_1$, $\d_2$) take some
special ($\d=-2(6-p)/(7-p)$, $\a_2 = 2$ or $\d_1 = -12/7$, $\d_2 =
-1/7$) values. This is expected since the black brane has two
parameters corresponding to its mass and the charge. This in turn
shows that the black branes are indeed of type 1 or intersecting
non-susy $p$-brane with $q$-brane solution of special kind.

(Intersecting) non-susy branes of type 1 has a natural
interpretation as (intersecting) brane-antibrane systems. Since
there are tachyons in the world-volume of these unstable systems,
they usually decay to BPS branes or Minkowski vacuum depending on
the initial charge of the system. This decay process (or the tachyon
condensation) can be described by relating the various SUGRA
parameters appearing in the solution semi-empirically to the
microscopic physical parameters of the system. But in this process
usually we do not encounter the formation of a regular formation. We
have demonstrated that this is because there exists two disjoint
processes, namely, the brane-antibrane annihilation process and the
black-brane related phase transition process in the parameter space
of non-susy branes. We have shown this by giving two concrete
examples, namely the D0/${\bar {\rm D}}$0 system and the
intersecting D4/${\bar {\rm D}}$4 and chargeless D0/${\bar {\rm
D}}$0 system where the explicit relations between the SUGRA
parameters and the microscopic physical parameters are known. The
brane-antibrane annihilation process where the underlying dynamics
is governed by the open string tachyon condensation is much better
understood and has been studied in specific cases in
refs.\cite{Lu:2004dp,Bai:2005jr}. The other process, i.e., the
regular horizon formation and the subsequent evolution or transition
to non-susy bubbles where the underlying dynamics is governed by the
possible closed string tachyon condensation except for the $p = 0$
case has been discussed by Horowitz and others
\cite{Horowitz:2005vp,Horowitz:2006mr}. The $p = 0$ case as well as
other issues as mentioned in the text are under current
investigation and we hope to come back to these in a future
publication.

\vspace{.5cm}

\noindent {\bf Acknowledgements}

\vspace{2pt}

JXL acknowledges support by grants from the Chinese Academy of
Sciences and grants from the NSF of China with Grant No: 90303002,
10588503 and 10535060.

\end{document}